\documentclass[sn-mathphys,Numbered]{sn-jnl}
\usepackage{graphicx}%
\usepackage{multirow}%
\usepackage{amsmath,amssymb,amsfonts}%
\usepackage{amsthm}%
\usepackage{mathrsfs}%
\usepackage[title]{appendix}%
\usepackage{xcolor}%
\usepackage{textcomp}%
\usepackage{manyfoot}%
\usepackage{booktabs}%
\usepackage{algorithm}%
\usepackage{algorithmicx}%
\usepackage{algpseudocode}%
\usepackage{listings}%
\usepackage[euler]{textgreek}
\usepackage{lineno}
\usepackage{soul}
\usepackage{xspace} 
\usepackage{booktabs,multirow}

\begin{document}
\newcommand{\rc}[1]{{\color{blue}{#1}}}
\newcommand{\tali}[1]{{\color{magenta}{#1 --Tali}}}
\newcommand{\Ricochet}{\textsc{Ricochet}\xspace}

\title[Article Title]{Modeling and characterization of TES-based detectors for the \Ricochet experiment}

\author*[1]{R. Chen}\email{ranchen2025@u.northwestern.edu}
\author[1]{E. Figueroa-Feliciano}
\author[1]{G. Bratrud}
\author[2]{C.L. Chang}
\author[3]{L. Chaplinsky}
\author[4]{E. Cudmore}
\author[5]{W. Van De Pontseele}
\author[5]{J. A. Formaggio}
\author[5]{P. Harrington}
\author[3]{S. A. Hertel}
\author[4]{Z. Hong}
\author[1]{K. T. Kennard}
\author[5]{M. Li}
\author[2]{M. Lisovenko}
\author[1]{L. O. Mateo}
\author[5]{D. W. Mayer}
\author[1]{V. Novati}
\author[3]{P.K. Patel}
\author[3]{H. D. Pinckney}
\author[1]{N. Raha}
\author[5]{F. C. Reyes}
\author[1]{A. Rodriguez}
\author[1]{B. Schmidt}
\author[5]{J. Stachurska}
\author[3]{C. Veihmeyer}
\author[2]{G. Wang}
\author[5]{L. Winslow}
\author[2]{V.G. Yefremenko}
\author[2]{J. Zhang}

\affil[1]{\orgdiv{Department of Physics and Astronomy}, \orgname{Northwestern University}, \orgaddress{\street{633 Clark St}, \city{Evanston}, \postcode{60208}, \state{IL}, \country{US}}}

\affil[2]{\orgname{Argonne National Laboratory}, \orgaddress{\street{9700 S Cass Ave}, \city{Lemont}, \postcode{60439}, \state{IL}, \country{US}}}

\affil[3]{\orgdiv{Department of Physics}, \orgname{University of Massachusetts Amherst}, \orgaddress{\street{710 N Pleasant St}, \city{Amherst}, \postcode{01003}, \state{MA}, \country{US}}}

\affil[4]{\orgdiv{Department of Physics}, \orgname{University of Toronto}, \orgaddress{\street{27 King's College Cir}, \city{Toronto}, \postcode{	M5R 0A3}, \state{ON}, \country{Canada}}}

\affil[5]{\orgdiv{Department of Physics}, \orgname{Massachusetts Institute of Technology}, \orgaddress{\street{77 Massachusetts Ave}, \city{Cambridge}, \postcode{02139}, \state{MA}, \country{Canada}}}

\abstract{Coherent elastic neutrino-nucleus scattering (CE\textnu NS) offers a valuable approach in searching for physics beyond the Standard Model. The \Ricochet experiment aims to perform a precision measurement of the CE\textnu NS spectrum at the Institut Laue–Langevin nuclear reactor with cryogenic solid-state detectors. The experiment plans to employ an array of cryogenic thermal detectors, each with a mass around 30\,g and an energy threshold of sub-100\,eV.
The array includes nine detectors read out by Transition-Edge Sensors (TES). These TES based detectors will also serve as demonstrators for future neutrino experiments with thousands of detectors.
In this article we present an update in the characterization and modeling of a prototype TES detector.}
\maketitle
\section{Introduction}\label{sec1}

In 1974, Daniel Freedman predicted that for sufficiently small momentum transfers, neutrinos can interact coherently with nuclei~\cite{Freedman:1974}.
The process, known as Coherent Elastic Neutrino(\textnu)-Nucleus Scattering (CE\textnu NS), evaded detection until 2017. The COHERENT experiment measured CE\textnu NS at the Spallation Neutron Source (SNS) with a neutrino energy of $\sim$30\,MeV~\cite{Akimov:2017}.
This measurement took advantage of the enhanced cross section with the coherent interaction, enabling a detection with kg-scale target mass rather than the traditional ton-scale neutrino detectors.

The \Ricochet experiment\cite{Augier2023, Augier2023_2} aims to measure the spectrum of CE\textnu NS with high-statistics and high-precision at the Institut Laue–Langevin (ILL) nuclear reactor in Grenoble, France.
The experiment is located 8.8\,m from the reactor core, which provides a high neutrino flux that is $\sim$1000 times higher than that from the SNS. However, the lower energy of neutrinos from the reactor requires a lower detector threshold ($\sim50$\,eV) to take advantage of this ﬂux.
To achieve such a low energy threshold, the \Ricochet experiment will employ cryogenic calorimeters. Two technologies---Neutron Transmutation Doped (NTD) thermistors and Transition Edge Sensors (TES)---will be used. The NTD-based detector array is called the Cryocube~\cite{Salagnac:2023, augier2023demonstration} and the TES-based detector array is named the Q-Array~\cite{AUGIER2023168765}. This paper focuses on the development of the TES-based detector and characterization of a prototype detector with a germanium target.

\section{Detector Architecture and Experiment Setup}

\subsection{Detector Architecture}
\label{Sec:Detector_Architecture}
A Q-Array style detector consists of two components~\cite{Chen:2023}: 
\begin{itemize}
\setlength{\itemindent}{1cm}
\item[1)] A target constituted by a crystal with a thin gold film deposited on the surface;
\item[2)]A modular thermal sensor equipped with a TES reading out the temperature excitation from the target.
\end{itemize}
In its quiescent state the detector target is held near the critical temperature of the TES. Particle interactions raise the temperature of the target. A 300-nm thick gold film is deposited with an electron beam evaporator onto the target to collect the energy released by the particle interaction through electron-phonon coupling. The absorbed energy is guided from the gold film to the TES sensor via a gold wire bond.
The other end of the TES sensor is connected to the thermal bath via an engineered thermal impedance. The thermal impedance, referred to as a ``meander", controls the heat flow through the TES. 
The TES and the meander are deposited on a $3 \times 3 \times 0.4$\,mm$^3$ silicon chip. The sensor modularity and separation from the target makes this detector architecture easily scalable and allows for flexibility in the target material choice.
The electrical signal generated by the TES is read out via two niobium traces which connect to a Superconducting Quantum Interference Device (SQUID) nearby the detector housing. 

%

\subsection{Experiment Setup}
We tested a germanium target with the modular TES-based sensor described in Sec.~\ref{Sec:Detector_Architecture}. The TES consists of an Al/Mn bi-layer with a critical temperature of 20\,mK, fabricated by Argonne National Laboratory. More details of the TES chip will be presented in a companion paper~\cite{Wang:2024}. The detector is housed in a light-tight copper holder. The germanium target is held by 8 sapphire spheres~\cite{Pinckney:2022} to isolate the target from copper holder, where in this case the copper holder functions as a thermal bath. The TES chip is glued to the housing using GE Varnish. The germanium target was activated using an Pu-Be source, resulting in the generation of an activated peak at 10.37\,keV from $^{71}\mathrm{Ge}$ decay.

The detector was operated in a Cryoconcept dilution refrigerator situated at the Northwestern Experimental Underground Site (NEXUS)~\cite{Ren:2021, Battaglieri:2017} at Fermilab (Batavia, IL). The cryostat was stabilized at a base temperature of 10\,mK. We used a single-stage DC SQUID to read out the TES.

\section{Complex Impedance Measurement and Modeling }
\label{Sec:Calib_Complex_Impedance}
\begin{figure}[htbp]
    \begin{minipage}{0.5\textwidth}
        \centering
        \includegraphics[width=1.0\linewidth]{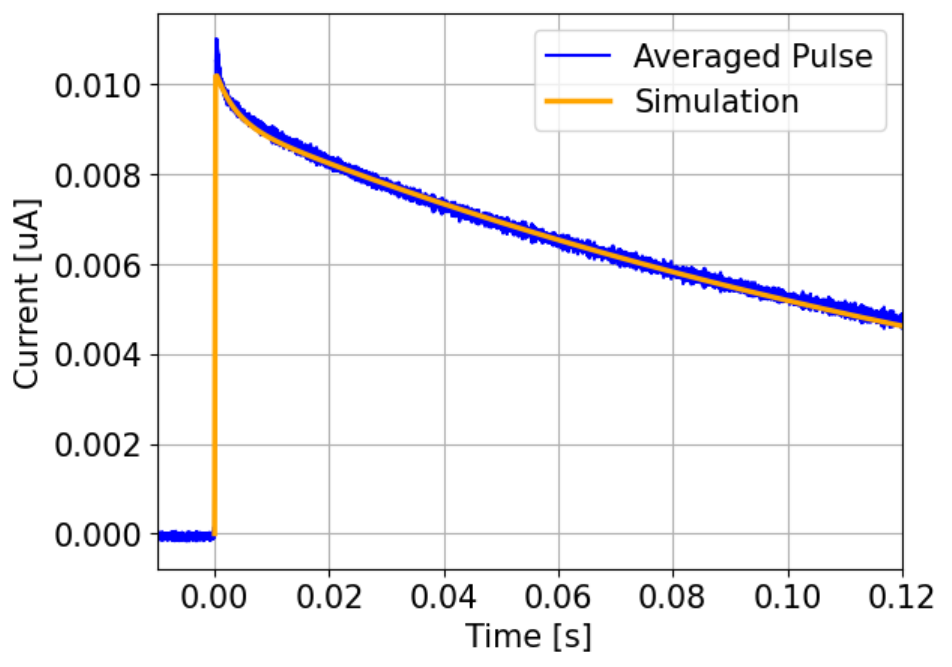}
    \end{minipage}
    \begin{minipage}{0.5\textwidth}
        \centering
        \includegraphics[width=1.0\linewidth]{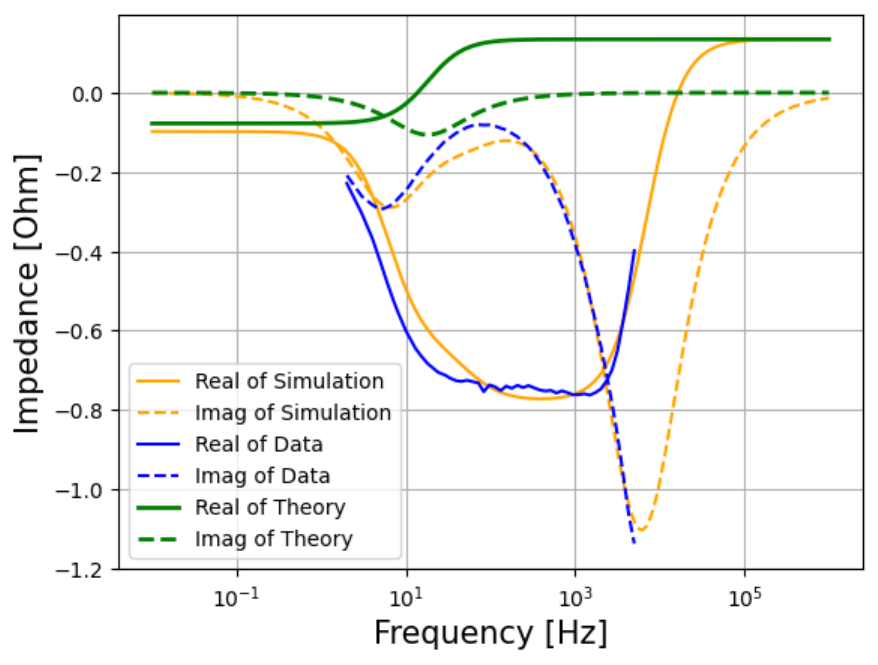}
    \end{minipage}
    \caption{\textit{Left panel}: average pulse (blue) from 10.37 keV and its simulation (orange).
    \textit{Right panel}: The data (blue) are compared to the simulation (orange) and the compled impedance for a TES-only device (green).}
    \label{fig7}
\end{figure}

The complex impedance of the detector was measured and leveraged to characterize the device. To measure the complex impedance, a signal generator injects sine waves with varying frequencies and amplitudes into the TES bias circuit~\cite{Hattori:2020}. Meanwhile, the readout circuit records the detector response to these injected signals.
The complex impedance was measured from 2\,Hz to 5\,kHz. Figure~\ref{fig7} shows the result of the complex impedance measurement with our detector and also for a TES-only device by theory. It illustrates that the complex impedance of our detector is significantly different from the theory.

\begin{figure}[htbp]
    \begin{minipage}{0.5\textwidth}
        \includegraphics[width=1.0\linewidth]{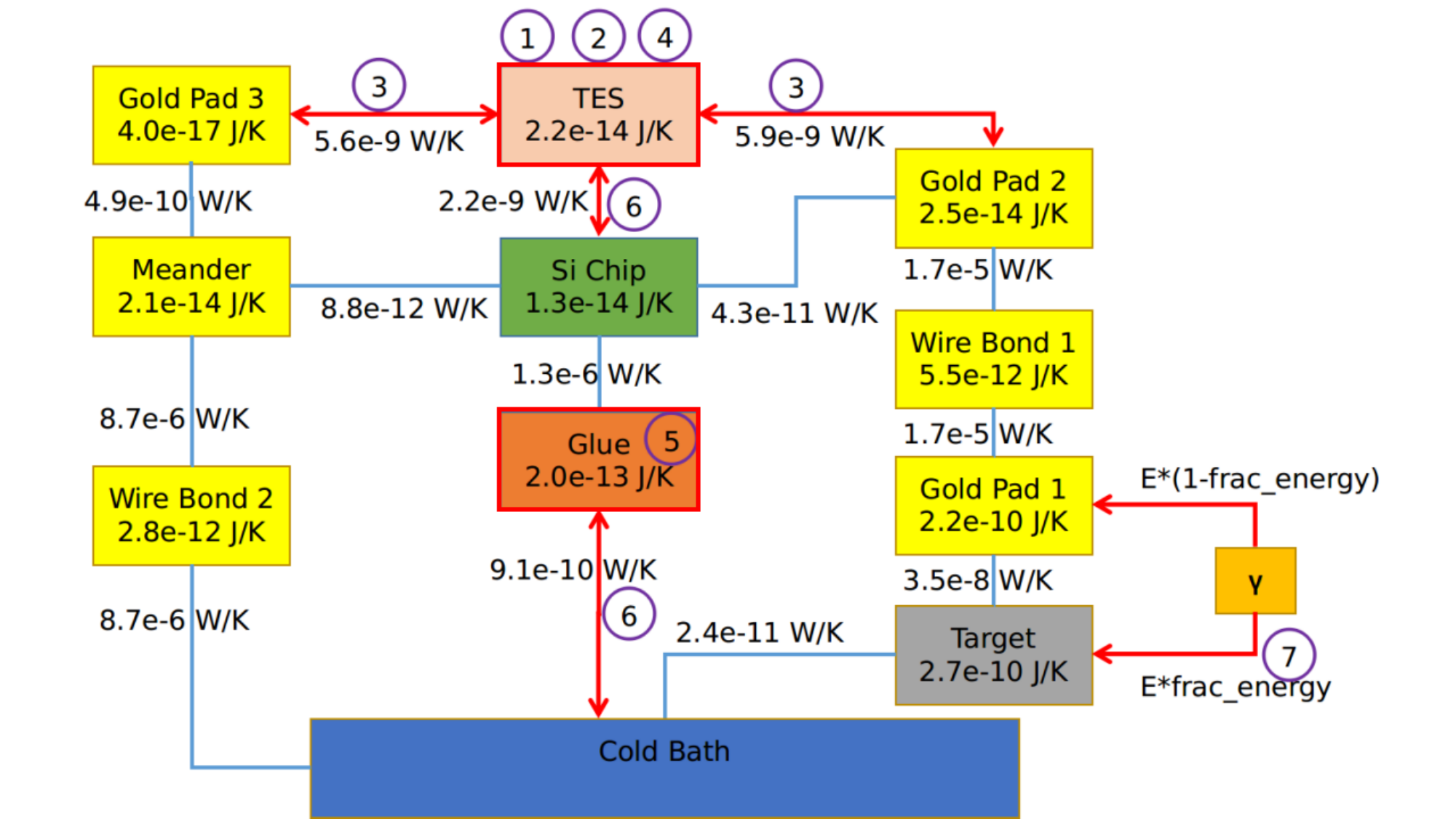}
    \end{minipage}
    \begin{minipage}{0.5\textwidth}
        \includegraphics[width=1.0\linewidth]{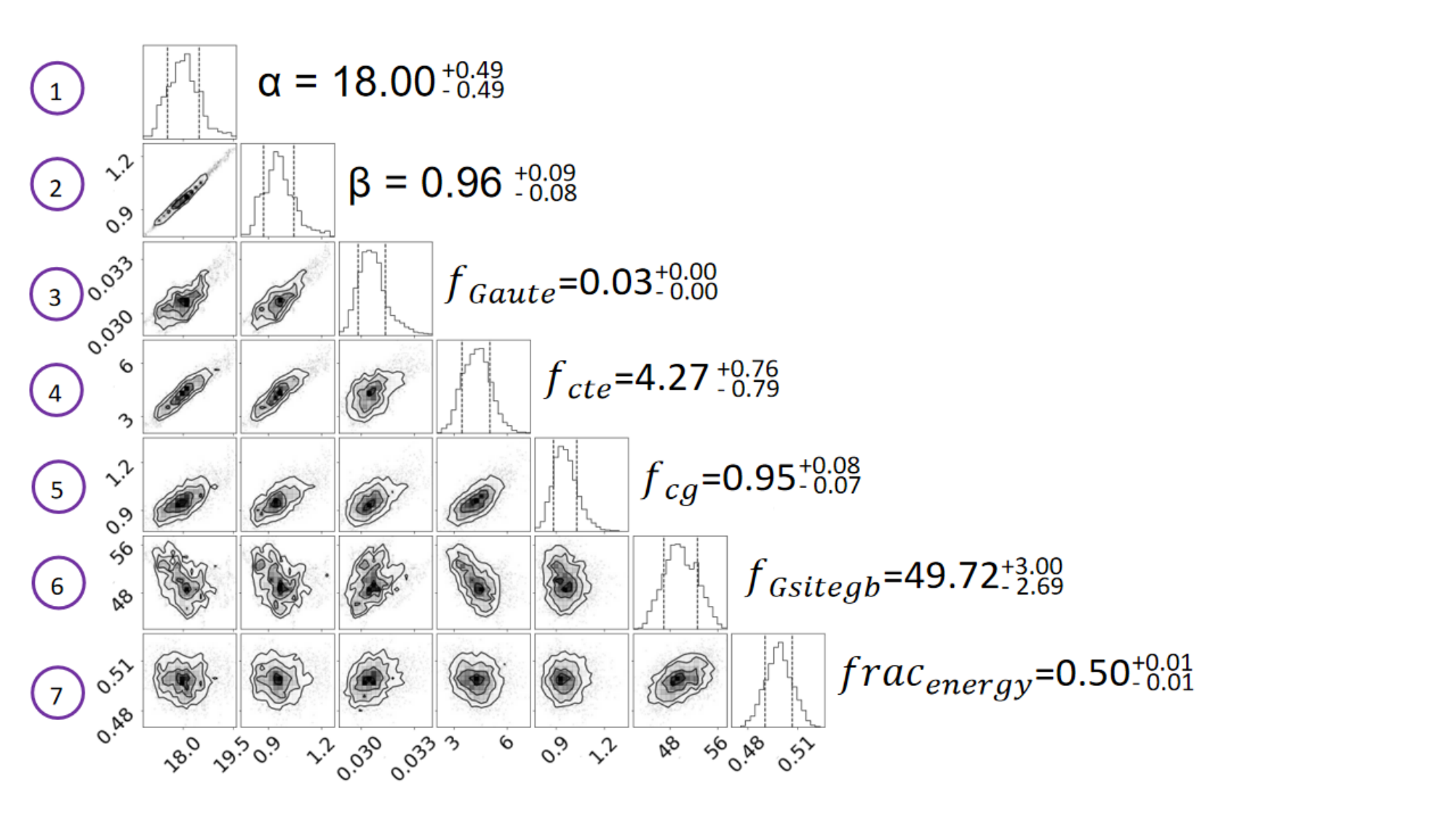}
    \end{minipage}
    \caption{\textit{Left panel}: Block-line diagram of the model. Each block represents a component with a heat capacity, while lines represent thermal conductances between these components. The components enclosed within a red box or marked with a red arrow were variated within the MCMC fitting process. All values are converged parameters from the MCMC fitting.
    \textit{Right panel}: Corner Plot from the MCMC Fitter. The meaning of each parameters is shown in Table~\ref{table1}. The histograms below each parameter show how the Markov chain distribution is spread out for that specific parameter. The density plots at the intersections of each pair of parameters reveal how the multidimensional distribution of the Markov chain is projected onto those two specific parameters.}
    \label{fig8}
    \end{figure}

In order to gain a better understanding of the detector's behavior, optimize its design and improve its future performance, we constructed a numerical model for simulating this specific type of detector \cite{Chen:2023, Figueroa:2006, Bastidon2018}. In the model, each component is simplified as a point with a heat capacity and zero thermal diffusion relaxation time, as shown in Fig.~\ref{fig8}. The thermal connections between components are described by a set of linearized thermal diffusion equations. 

This work aimed to model both the static and dynamic response of the detector by reproducing the complex impedance results and the average pulse and allows us to estimate:
\begin{itemize}
\setlength{\itemindent}{1cm}
\item[1)] Heat capacities of each component;
\item[2)] Thermal conductance between each component;
\item[3)] Temperature sensitivity and current sensitivity of the TES~\cite{Irwin:2005}.
\end{itemize}
A Markov Chain Monte Carlo (MCMC) was used to fit the complex impedance data. Parameters selected for the MCMC optimization are highlighted with a red box or arrow in Fig.~\ref{fig8} (left).

\begin{table}[t]
\begin{tabular}{llccc} 
 \toprule
 Categories & Parameters &units & MCMC  & MCMC \\%
   & & &  initialization &  results\\
 \midrule\\[-.9em]
  \midrule\\[-.9em]
 \multirow{2}{*}{\shortstack[l]{TES\\Parameters}}
 & Temperature sensitivity of TES & No Unit & 9 & 18\\
 & Current sensitivity of TES & No Unit & 1.00 & 0.96\\[.5em]
  \midrule\\[-.9em]
 \multirow{9}{*}{\shortstack[l]{Heat\\Capacities}}
 & Germanium Target & J/K & $2.7\times10^{-10}$ & $2.7\times10^{-10}$\\
 & Au Pad 1 & J/K & $2.2\times10^{-10}$ & $2.2\times10^{-10}$ \\
 & Wire Bond 1 & J/K & $5.5\times10^{-12}$  & $5.5\times10^{-12}$ \\
 & Au Pad 2 & J/K & $2.5\times10^{-14}$  & $2.5\times10^{-14}$ \\
 & TES & J/K & $5.1\times10^{-15}$  & $2.2\times10^{-14}$ \\
 & Si Chip & J/K & $1.3\times10^{-14}$  & $1.3\times10^{-14}$ \\
 & Meander & J/K & $2.1\times10^{-14}$  & $2.1\times10^{-14}$ \\
 & Wire Bond 2 & J/K & $2.8\times10^{-12}$  & $2.8\times10^{-12}$ \\
 & Glue & J/K & $2.1\times10^{-13}$  & $2.0\times10^{-13}$ \\[.5em]
 \midrule\\[-.9em]
\multirow{14}{*}{\shortstack[l]{Thermal\\Conductances}}
& Between Target and Cold Bath & W/K & $2.4\times10^{-11}$ & $2.4\times10^{-11}$ \\
& Between Target and Au Pad 1 & W/K & $3.5\times10^{-8}$  & $3.5\times10^{-8}$ \\
& Between Au Pad 1 and Wire Bond 1 & W/K & $1.7\times10^{-5}$  & $1.7\times10^{-5}$ \\
& Between Wire Bond 1 and Au Pad 2 & W/K & $1.7\times10^{-5}$ & $1.7\times10^{-5}$\\
& Between Au Pad 2 and TES  & W/K & $1.9\times10^{-7}$  & $5.9\times10^{-9}$ \\
& Between Au Pad 3 and TES  & W/K & $1.9\times10^{-7}$  & $5.9\times10^{-9}$ \\
& Between Au Pad 2 and Si Chip  & W/K & $4.3\times10^{-11}$  & $4.3\times10^{-11}$ \\
& Between TES and Si Chip  & W/K & $4.4\times10^{-11}$  & $2.2\times10^{-9}$ \\
& Through Gold Pad 3 and Meander  & W/K & $4.9\times10^{-10}$  & $4.9\times10^{-10}$ \\
& Between Meander and Si Chip  & W/K & $8.8\times10^{-12}$  & $8.8\times10^{-12}$ \\
& Between Si Chip and Glue  & W/K & $1.3\times10^{-6}$  & $1.3\times10^{-6}$ \\
& Between Glue to Cold Bath  & W/K & $1.8\times10^{-11}$  & $9.1\times10^{-10}$ \\
& Between Meander and Wire Bond 2 & W/K & $8.7\times10^{-6}$ & $8.7\times10^{-6}$ \\
& Between Wire Bond 2 and Cold Bath & W/K & $8.7\times10^{-6}$ & $8.7\times10^{-6}$ \\[.5em]
 \midrule\\[-.9em]
\multirow{1}{*}{\shortstack[l]{Simulation\\Parameter}}
& Fraction of Energy into Target & No Unit & 100\% & 50\% \\[.5em]
\bottomrule
\end{tabular}
\caption{Initial parameter values and final fit values for the heat capacities, thermal conductances, TES parameters and energy fraction used in the MCMC fit.}
\label{tab2}
\end{table}

The MCMC method is implemented through a python package called ``PyMC3"~\cite{Salvatier:2016} which uses gradient-based MCMC algorithms.\
Figure~\ref{fig7} shows the complex impedance data and the average pulse (blue), and its MCMC fit to the simulation (orange). Parameters from the MCMC fit are reported in Table~\ref{tab2}.
Figure~\ref{fig8} (right) shows the corner plot of the MCMC fit which illustrates the distribution of each parameter in the Markov chain and their correlation with the other  parameters. A definition of parameters in the corner plot can be found in Table~\ref{table1}.

Despite being a thermal detector, athermal processes can still occur and impact the pulse shape. To simulate this phenomenon in the model, instead of injecting the total energy from an event directly into the target, a fraction of the energy is injected into gold pad 1. The thermal impedance between this pad and the TES is smaller and can be used to simulate the effects of faster thermalization to the gold through athermal processes. The fraction of energy injected into the gold pad 1 is defined as $\mathrm{frac_{energy}}$, as shown in the corner plot Fig.~\ref{fig8} (right).

A discussion of results follows:
\begin{itemize}
\setlength{\itemindent}{1cm}
\item[1)] The value $\mathrm{G_{aute}}$ is the thermal conductance between gold pad 2 and 3, and the TES. The difficulty in accurately estimating the thermal conductance through the interface between nanometer scale thin films
has resulted in a large uncertainty. To avoid over-constraining the fit, we have estimated the thermal conductance through the gold pad, which sets a fairly high upper limit for $\mathrm{G_{aute}}$, and assume a flat prior down to zero conductance. The corner plot shows that the fit prefers a value 3 \% of the first estimation. A lower conductance indicates a slow response time of the detector and a lower the energy efficiency because some energy will bypass the TES. In future experiments, we will employ a different chip which has a larger overlap area between gold pad and TES to improve this conductance.
\item[2)] The factor $\mathrm{f_{Gsitegb}}$ controls the thermal conductance from TES through glue to the cold bath. The value of 50 from the MCMC fitting indicates a large thermal conductance through the glue which could allow energy to enter the thermal bath without passing through the TES. This detrimental effect will be mittigated by new mounting strategies, such as clamping the chip mechanically, will be tested to reduce this effect.
\end{itemize}

\begin{table}[htbp!]
\begin{tabular}{ccl}
 \toprule
Ref. \# in            & Ref. in                  &\\
Fig.~\ref{fig8}   & Fig.~\ref{fig8}  &  Description   \\
(left)  & (right)  &     \\
 \midrule
  1 & \textalpha & \textalpha, Temperature sensitivity of TES \\ 
  2 & \textbeta & \textbeta, Current sensitivity of TES \\
  3 & $\mathrm{f_{Gaute}}$  & Factor for Therm. Cond. between Gold and TES\\ 
  4 & $\mathrm{f_{cte}}$  & Factor for heat capacity of TES\\
  5 & $\mathrm{f_{cg}}$  & Factor for heat capacity of glue\\
  6 & $\mathrm{f_{Gsitegb}}$& Factor for Therm. Cond. from TES to the cold bath\\
  7 & $\mathrm{frac_{energy}}$&Fraction of energy injected into target, rest injects to the gold pad 1\\
 \bottomrule
\end{tabular}
\caption{Definition of the parameters varied in MCMC fitting. The heat capacities and thermal conductances (Therm. Cond.) are multiplied by a factor.}
\label{table1} 
\end{table}

\section{Conclusion and Plan}
A germanium detector equipped with a modular TES-based sensor was tested and characterized at the NEXUS test facility. Currently we are suffering from a vibrations induced noise by the pulse tube so that the energy resolution is not ideal so we focused on the complex impedance measurement and modeling.

The successful concurrent fitting of pulse shape and complex impedance measurements have enabled us to gain insights into the heat capacities and thermal conductance of the system as well as identify key limitations affecting performance. For example, the thermal conductance from the target to TES is lower than the original estimation which may limit the energy efficiency.
This limit will be tackled with a newer version of the modular sensor design, which is capable of improving thermal diffusion along the TES thanks to a larger overlap area between gold and TES, thereby reducing the decay time constant and increase the energy efficiency. These ongoing efforts are geared towards further enhancing the performance and capabilities of the detector system.

\section{Acknowledgements}
This work is supported in part by NSF grant PHY-2209585, the Connaught Fund at University of Toronto, and the Canada First Research Excellence Fund through the Arthur B. McDonald Canadian Astroparticle Physics Research Institute. This research was enabled in part by support provided by SciNet (www.scinethpc.ca) and the Digital Research Alliance of Canada (alliancecan.ca). This work is part of R\&D for \Ricochet experiment and we would like to thank all members of \Ricochet.

This work made use of the NUFAB facility of Northwestern University’s NUANCE Center, which has received support from the SHyNE Resource (NSF ECCS-2025633), the IIN, and Northwestern’s MRSEC program (NSF DMR-1720139).

Work at Argonne National Lab, including work performed at the Center for Nanoscale Materials, a U.S. Department of Energy Office of Science User Facility, is supported by the U.S. Department of Energy, Office of Science, Office of High Energy Physics and Office of Basic Energy Sciences, under Contract No. DE-AC02-06CH11357.

A portion of the work carried out at MIT was supported by a U.S. Department of Energy (DOE) DOE QuantISED award DE-SC0020181, the NSF under Grant PHY-2110569, and the Heising-Simons Foundation, United States.

\section*{Data Availability}

The datasets analysed during the current study are available from the corresponding author on reasonable request.

\bibliography{main}

\end{document}